\documentclass{ws-procs9x6}

\begin{document}

\title{Evidence for Neutrino Mass: \\ A Decade of Discovery}

\author{Karsten~M.~Heeger \footnote{\uppercase{T}his work was supported by the \uppercase{D}irector, \uppercase{O}ffice of \uppercase{S}cience, \uppercase{H}igh \uppercase{E}nergy \uppercase{P}hysics, \uppercase{U}.\uppercase{S}.~\uppercase{D}epartment of \uppercase{E}nergy under contract no. \uppercase{DE}-\uppercase{AC}03-76\uppercase{SF}00098.}}

\address{Lawrence Berkeley National Laboratory \\
Physics Division, MS50R5008 \\ 
Berkeley, CA 94720, USA\\ 
E-mail: kmheeger@lbl.gov}

\maketitle

\abstracts{
Neutrino mass and mixing are amongst the major discoveries of recent years.  From the observation of flavor change in solar and atmospheric neutrino experiments to the measurements of neutrino mixing with terrestrial neutrinos, recent experiments have provided consistent and compelling evidence for the mixing of massive neutrinos. The discoveries at Super-Kamiokande, SNO, and KamLAND have solved the long-standing solar neutrino problem and demand that we make the first significant revision of the Standard Model in decades. Searches for neutrinoless double-beta decay  probe the particle nature of neutrinos and continue to place limits on the effective mass of the neutrino. Possible signs of neutrinoless double-beta decay will stimulate neutrino mass searches in the next decade and beyond. I review the recent discoveries in neutrino physics and the current evidence for massive neutrinos.
}


\section{Neutrinos Within the Standard Model}

In 1930 Pauli postulated the neutrino as a ``desperate remedy'' to the energy crisis of the time - the continuous energy spectrum of electrons emitted in nuclear $\beta$-decay. With the postulate of a new particle Pauli could account for the continuous spectrum. He assumed that nuclear $\beta$-decay emits a neutron together with an electron in such a way that the sum of the energies is constant. Sensitive measurements of the energy and momentum of $\beta$-decay electrons and the recoiling nuclei in cloud chambers indicated that substantial quantities of energy and momentum were missing. These experiments left little doubt that a third particle had to be involved. As early as 1932 Enrico Fermi provided a theoretical framework for $\beta$-decay which included the neutrino but it took another 25 years before the neutrino was detected experimentally. In 1957 Frederick Reines and Clyde Cowan made the first observation of the free antineutrino through the inverse $\beta$-reaction $\overline{\nu}_{e} + \rm{p} \rightarrow e^+ + \rm{n}$ utilizing the flux of $\overline{\nu}$ from the Savannah River nuclear reactor \cite{reines}. The muon neutrino was finally detected by Schwartz, Lederman, and Steinberger in 1961 \cite{steinberger}. Neutrinos from pion and kaon decays with energies of hundreds of MeV to several GeV were detected in a 10-ton spark chamber built from aluminum plates that provided distinct signals for the showering of electrons and the tracks of muons generated by neutrinos. The excess of muons produced in the chamber provided evidence that the neutrino produced in the decay of $\pi \rightarrow \mu + \nu$ was distinct from those produced in $\beta$-decays. The total number of light neutrino types, $N_{\nu}$, has been deduced from the studies of Z production in $e^+e^-$ collisions. Assuming that the invisible partial decay width is due to light neutrino species with partial width $\Gamma_{\nu}$ the number of active, light neutrino species is given by

\begin{eqnarray}
N_{\nu} = \frac{\Gamma_{inv}}{\Gamma_{l}}\left( \frac{\Gamma_{l}}{\Gamma_{\nu}}\right)
\end{eqnarray}

The LEP and SLC lineshape measurement of Z bosons left little doubt about the existence of the $\nu_{\tau}$. In 2001 the existence of $\nu_{\tau}$ was confirmed by the Fermilab DONUT experiment \cite{donut}. Seven decades after Pauli's postulate it was experimentally  established that there are three neutrinos associated with the flavor of their leptonic interaction. In the absence of any other insight the neutrino was assumed to be massless, an ad-hoc assumption in the Standard Model of particle physics.

\section{Birth of the Solar Neutrino Problem}

\begin{figure}[ht]
\centerline{\epsfxsize=3.5in\epsfbox{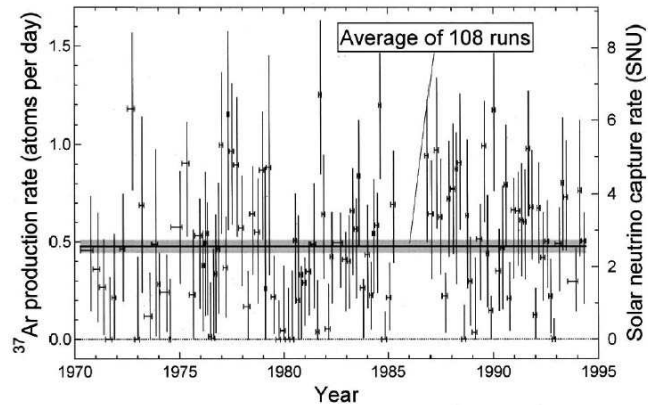}}   
\caption[]{Measured solar neutrino capture rate in Ray Davis' chlorine experiment at Homestake between 1970 and 1995. The observed average solar neutrino flux was 2.56$\pm$0.16$\pm$0.16 SNU, about a third of the current solar model prediction of 7.6$^{+1.3}_{-1.1}$ SNU. The neutrino  capture rate is given in Solar Neutrino Units (SNU). One SNU is equivalent to  10$^{-36}$s$^{-1}$ interactions per nucleus. Figure from \cite{davisnobel}. \label{homestakedata}}
\end{figure}

Around the same time Pauli postulated the neutrino, Bethe and Critchfield proposed pp fusion, $\rm{p}+\rm{p} \rightarrow$~$^2\rm{H} + e^+ + \nu_{e}$, as the mechanism for solar energy generation. The solar nuclear reactions fuse protons into helium and release neutrinos with energies of up to 15~MeV. As early as 1946 and 1949, Bruno Pontecorvo and Luis Alvarez proposed independently neutrino detection via $\nu_{e}$ capture on chlorine through $^{37}\rm{Cl} + \nu_{e} \rightarrow$~$^{37}\rm{Ar}+ e^-$.  Using this idea Ray Davis built a chlorine detector in the 1960's to detect neutrinos from the Sun and {\em``to see into the interior of a star and thus verify directly the hypothesis of nuclear energy generation in stars'}'.  The efforts of Ray Davis and John Bahcall in the measurement of the solar neutrino flux and the development of solar models and the prediction of the solar neutrino flux resulted in the birth of neutrino astrophysics. In his experiment Ray Davis measured a significantly lower flux of solar neutrinos than predicted by current solar models. The results 
from this first solar neutrino experiment are shown in Figure~\ref{homestakedata}.

For more than 30 years now, experiments have observed neutrinos produced in the thermonuclear fusion reactions which power the Sun.  The solar neutrino flux has been measured through the charged-current ($^{37}\rm{Cl} + \nu_{e} \rightarrow$~$^{37}\rm{Ar} + e^-$, $^{71}\rm{Ga} + \nu_{e} \rightarrow$~$^{71}\rm{Ge} + e^-$) or elastic scattering ($\nu_{x} +e^- \rightarrow \nu_{x} + e^-$) channels. Data from these solar neutrino experiments were found to be incompatible with the predictions of solar models. More precisely, the flux of neutrinos detected on Earth was less than expected, and the measured relative intensities of the neutrino sources in the Sun were incompatible with those predicted by solar models.

A variety of hypotheses including neutrino decay were postulated to explain the discrepancy between solar model expectations and the apparent deficit of solar neutrinos detected on Earth. As early as 1969, Bruno Pontecorvo proposed that neutrinos might oscillate between the electron and muon flavor, the only states known at the time \cite{pontecorvo}. Oscillations can occur if the physical neutrinos consist of a superposition of mass states. If neutrinos are massive an initially pure flavor state changes as the neutrino propagates. Neutrino mass and flavor mixing are not features of the Standard Model of particle physics and neutrino flavor change through oscillation requires the existence of massive neutrinos. For two neutrino flavors the survival probability of neutrinos in vacuum is given by 

\begin{eqnarray}
P(\nu_{l} \rightarrow \nu_{k}) = \sin^22\theta\sin^2\left( 1.27 \Delta m^2 \frac{L}{E_{\nu}} \right)
\end{eqnarray}

where $\Delta m^2 = \left| m^2_{2}-m^2_{1} \right|$ is given in eV$^2$,  $L$ in km and $E_{\nu}$ in GeV. The neutrino survival probability $P_{l \rightarrow k}$ depends on ratio of the distance traveled over the energy of the neutrino $L/E$ and the mixing angle $\theta$ and the mass splitting $\Delta m^2$. $L/E$ is usually determined by experiments while $\theta$ and $\Delta m^2$ are fundamental parameters of nature. 

\section{The Atmospheric Neutrino Anomaly}

Strong indication for neutrino oscillation first came from the observation of atmospheric neutrinos. These neutrinos are the decay products of hadronic showers produced by cosmic ray interactions in the atmosphere. The pion production in the atmosphere determines the flux of atmospheric neutrinos incident on Earth. Around 1~GeV, where the product of flux and neutrino charged-current interactions cross-section reaches a maximum, the atmospheric neutrino flux is about 1 cm$^{-2}$s$^{-1}$. Atmospheric neutrinos span an energy range from $\sim$ 0.5-5 GeV. The path length of downgoing and upward going atmospheric neutrinos varies from 10-10,000~km and provides an opportunity for oscillation studies over a wide range of distances.

\begin{figure}[ht]
\centerline{\epsfxsize=3.5in\epsfbox{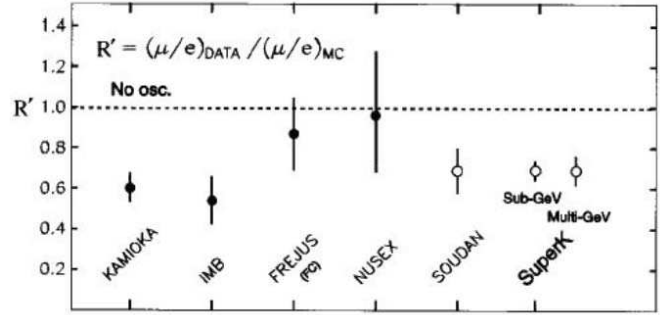}}   
\caption{Measurements of the double ratio R$_{atm}$=$\frac{(\nu_{\mu}+\overline{\nu}_{\mu})/(\nu_{e}+\overline{\nu}_{e})_{Data}}{(\nu_{\mu}+\overline{\nu}_{\mu})/(\nu_{e}+\overline{\nu}_{e})_{MC}}$ in atmospheric neutrino experiments. The ratio denotes the ratio of the number of $\mu$-like to $e$-like neutrino interactions. R$_{atm}$ estimates the atmospheric neutrino flavor ratio and is expected to be 1. \label{atmdata}}
\end{figure}

If all pions and muons decay we expect to observe about two $\mu$-like neutrinos for each $e$-like neutrino. The ratio $\nu_{e}/\overline{\nu}_{e}$ is expected to be close to $\mu^{+}/\mu^{-}$, about 1.2 at 1~GeV. Several effects including the muon decay length at high energies above 2.5~GeV, the muon energy loss, the geomagnetic cut-off and the variations of the cosmic ray flux with the solar cycle affect this prediction. Figure~\ref{atmdata} shows the results from various experiments that measured the total atmospheric neutrino flux. The ``ratio of ratio'' R$_{atm}$ is used to compare the results of various experiments. A number of experiments found a statistically low value of R$_{atm}$. 

\begin{figure}[ht]
\centerline{\epsfxsize=4.6in\epsfbox{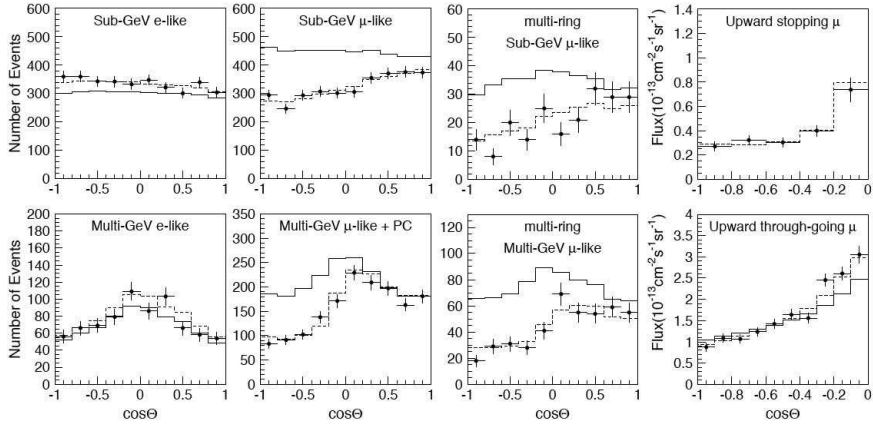}}   
\caption[]{Zenith-angle distribution for fully-contained single-ring $e$-like and $\mu$-like events, multi-ring $\mu$-like events, partially contained events, and upward-going muons. The points are the data and the solid lines show the Monte Carlo events without neutrino oscillation. The dashed lines show the best-fit expectations for $\nu_{\mu} \rightarrow \nu_{\tau}$ oscillation. Figure from \cite{SKatm}. \label{SKzenith}}
\end{figure}

The Super-Kamiokande experiment has measured the up-down asymmetry of the the atmospheric $\nu_{\mu}$ flux as well as the zenith angle distribution \cite{SKatm}. Figure~\ref{SKzenith} shows the experimental result. Only a zenith-angle dependent transformation of neutrino flavors can explain these measurements. Upward going  neutrinos traverse a much longer distance and have time to oscillate whereas downward going neutrinos do not. For electron neutrinos the event rate is independent of direction and the solid angle. The hypothesis of $\nu_{\mu} \rightarrow \nu_{\tau}$ oscillations fits well the angular distribution of the atmospheric neutrino flux. In the oscillation model, the mixing angle $\theta_{23}$ is found to be near maximal and the separation of mass states is about $\Delta m^2_{atm} \sim 2 \times 10^{-3}~{\rm eV^2}$. A recent analysis of the $L/E$ distribution of the data excludes neutrino decay and decoherence as possible explanations \cite{SKatm}. Besides oscillation, no other consistent particle physics explanation has been proposed to explain the atmospheric neutrino result. 

\section{Neutrino Flavor Change in Solar Neutrinos}

Solar neutrinos are produced in the light element fusion reactions that power the Sun. Using input from astronomical observation, nuclear physics, and astrophysics, models have been developed that allow us to make detailed predictions of the life cycle of stars and  their energy generation. Solar models trace the evolution of the Sun over the past 4.7 billion years of main sequence burning, thereby predicting the present-day temperature and composition profile of the solar core. It is believed that thermonuclear reaction chains generate the solar energy. Standard solar models (SSM) predict that over 98\% of the solar energy is produced from the pp-chain conversion of four protons into $^4$He, $4\rm{p} \rightarrow$ $^4\rm{He} + 2e^+ + 2\nu_{e}$, while the proton burning through the CNO cycle contributes the remaining 2\%. According to standard solar models only 0.01\% of the total solar neutrino emission is produced in the $\beta$-decay of $^8$B $\rightarrow$ $^8$Be +e$^+$+$\nu_{e}$. Solar models are constrained to produce today's solar radius, mass, and luminosity. The predictions of these models are in good agreement with recent observations from helioseismology and other observables.  

Over the past 30 years the flux of electron neutrinos from the Sun has been detected and measured in a number of experiments using a variety of experimental techniques. All solar neutrino experiments found indications for a suppression of the solar neutrino flux. With different detection thresholds the experiments have provided information across the entire solar neutrino spectrum from sub-MeV to about 15~MeV. In all cases the solar neutrino flux measurements  fall significantly below the predictions of the standard solar models. By the mid-1990's the data were beginning to suggest that one could not even in principle adjust solar models sufficiently to account for the effects. A model-independent analysis of the available data showed that no change in the solar models can completely account for the discrepancy between data and the energy-dependent solar neutrino flux predictions \cite{heegerrobertson}. If the experimental uncertainties are correctly estimated and the Sun is generating energy by light-element fusion in quasistatic equilibrium, the probability of a solution to the solar neutrino problem within the minimal Standard Model of particle physics is less than 2\%. Novel neutrino properties seemed to be called for. The long-standing Solar Neutrino Problem indicated that either solar models are incorrect and do not predict correctly the neutrino production and emission from the Sun or solar neutrinos undergo a flavor-changing transformation on their way from the Sun to Earth, and the electron solar neutrino flux detected in all first-generation solar neutrino experiments was only a component of the total solar neutrino flux.

With the recent measurements of the Sudbury Neutrino Observatory (SNO) \cite{SNONIM} it has finally become possible to test solar model predictions and the particle properties of neutrinos independently. With D$_2$O as its target medium the SNO detector is uniquely suited to make a simultaneous measurement of the solar $\nu_{e}$ flux and the total flux of all active $^8$B neutrinos. In SNO, solar $^8$B neutrinos interact with deuterium in three different reactions: The charged-current interaction of electron neutrinos with deuterium (CC), the neutral-current dissociation of deuterium though the interaction with active neutrino flavors, and the  elastic scattering off electrons. Only the charged-current reaction is exclusively sensitive to $\nu_{e}$. \\

\indent\indent (CC) $\nu_{e} + d \rightarrow p+p+e^-$ \\
\indent\indent (NC) $\nu_{x} + d \rightarrow p+n+e^-$ \\
\indent\indent (ES) $\nu_{x} + e^- \rightarrow nu_{x}+e^-$ \\

The sensitivity of SNO to the neutral current channel, the total flux of active solar neutrinos, allows it to make several key measurements: Comparing the NC to CC interaction rate SNO can test directly for neutrino flavor change independent of any solar model predictions. The measurement of the total flux of solar $^{8}$B neutrinos provides a good test of neutrino flux predictions in solar models. The diurnal time dependence and distortions in the neutrino spectrum are direct signatures of neutrino oscillation. 

Using first pure D$_2$O and then heavy water with dissolved NaCl to increase the neutron capture energy and efficiency in the NC interaction channel, SNO has measured the total solar neutrino flux \cite{SNOsalt}
\begin{eqnarray}
\phi^{SNO}_{total}=5.21 \pm 0.27 {\rm (stat)} \pm 0.38 {\rm (syst)}   \times 10^6 {\rm cm^{-2}s^{-1}}
\end{eqnarray}

\begin{figure}[ht]
\centerline{\epsfxsize=3.2in\epsfbox{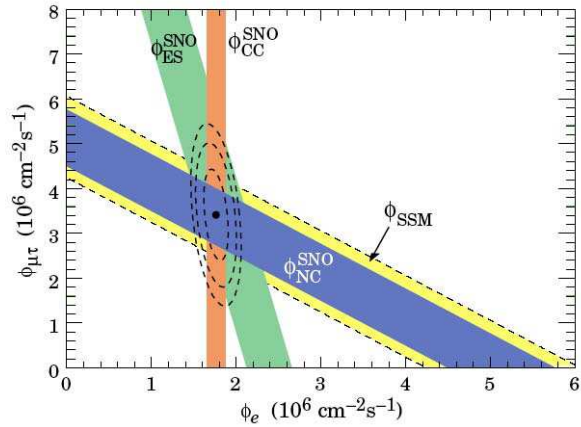}}   
\caption[]{Measured $^8$B solar $\nu_{\mu}$  and $\nu_{\tau}$ flux, $\phi_{\mu,\tau}$, versus the observed $\nu_{e}$ flux, $\phi_{e}$, deduced from the NC, ES, and CC neutrino interaction rates in SNO. The dashed lines show the total $^8$B neutrino flux as predicted by standard solar models. The bands intersect at the fit values for $\phi_{e}$ and $\phi_{\mu,\tau}$. This illustrates the $\nu_{e} \rightarrow \nu_{\mu,\tau}$ transformation of  solar $^8$B solar neutrinos. Figure from \cite{PDG}. \label{SNOd2o}}
\end{figure}

The interaction rates in the NC, CC, and
ES channels are determined from a statistical separation of events using the angular distribution, the event isotropy, and characteristic detector distributions. This measurement of the solar $^8$B $\nu$ flux is in excellent agreement with previous measurements and standard solar models. The ratio of the solar electron neutrino flux to the total flux of active $^8$B neutrinos indicates a clear ``deficit'' of solar electron neutrinos:
\begin{eqnarray}
\frac{\phi^{SNO}_{CC}}{\phi^{SNO}_{NC}}=0.306 \pm 0.026 \rm{(stat)} \pm 0.024 \rm{(syst)}
\end{eqnarray}
Figure~\ref{SNOd2o} shows a summary of the SNO solar neutrino flux measurements from the D$_2$O phase of the experiment \cite{SNOd2o}. 

\begin{figure}[ht]
\centerline{\epsfxsize=3.2in\epsfbox{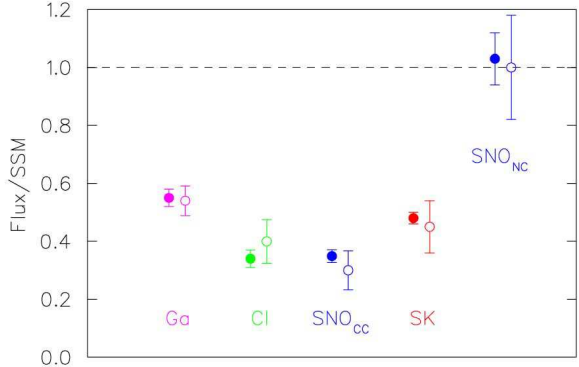}}   
\caption[]{Ratio of the measured solar neutrino flux to the predicted flux (in the absence of $\nu$ oscillations) for various experiments.  The experimental data (filled circles) and the best-fit predictions from the $\nu$ oscillation hypothesis (open circles) are in good agreement. Figure from \cite{mckeown2004}.\label{mckeown2004}}
\end{figure}

The $\phi_{e}$ and $\phi_{\mu,\tau}$ measurements of SNO alone provide direct evidence for the flavor change of solar neutrinos. Together with other solar neutrino experiments, the available experimental data probe different regions of the solar neutrino energy spectrum and  test the energy-dependent oscillation effect of solar neutrinos. Solar neutrinos pass through dense solar matter before they escape from the surface of the Sun and travel to Earth. The interaction of neutrinos with matter in the Sun and Earth creates an additional effective potential for electron neutrinos and enhances the oscillation probability of $\nu_{e}$ by shifting the energy of the states through the so-called MSW effect \cite{MSW}. The model of matter-enhanced neutrino oscillation provides an excellent description of the available solar neutrino data, as shown in Figure~\ref{mckeown2004}.

\section{Signatures of Neutrino Oscillation in Reactor Experiments}

Reactor neutrino experiments have played an important role in the history of neutrino physics. From the first direct detection of the antineutrino by Frederick Reines and Clyde Cowan \cite{reines} to the searches for neutrino oscillation and neutrino magnetic moment. For the past five decades nuclear power plants have been used as sources for low-energy studies with electron antineutrinos. Fission reactions in $^{238}$U,$^{239}$Pu, $^{241}$Pu,  $^{235}$U, and other isotopes produce $\sim 6$ $\overline{\nu}_{e}$ per fission with an energy release of about 200~MeV per fission. On average nuclear reactors produce $\sim 2 \times 10^{20}~\overline{\nu}_{e}$ per GW$_{th}$-sec with energies up to $\sim 8$~MeV and an average energy of about 4~MeV.  

The observation of neutrino flavor change with large mixing at a mass splitting of $\Delta m^2 \sim 7.1 \times 10^{-5}~{\rm eV^2}$ in solar neutrinos suggests 
that neutrinos or antineutrinos undergo oscillations in vacuum with a baseline of $\mathcal{O}$(100~km). The Kamioka Liquid Scintillator Antineutrino Detector (KamLAND) located in the Kamioka underground laboratory in Japan is uniquely suited to measure the flux of reactor $\overline{\nu}_{e}$. With 1000~t of liquid scintillator detector KamLAND measures the interaction rate and energy spectrum of $\overline{\nu}_{e}$ using the coincidence signal of the $e^+$ annihilation and the neutron capture in the inverse $\beta$-reaction $\overline{\nu}_{e} + \rm{p} \rightarrow e^+ + \rm{n}$. About 95\% of the $\overline{\nu}_{e}$ flux at KamLAND comes from commercial power plants in Japan. The flux-averaged mean baseline is about 180~km. 

In 2003, KamLAND made the first direct observation of reactor $\overline{\nu}_{e}$ disappearance. Based on an exposure of 162 kt-yr KamLAND observed 54 events above 2.6~MeV compared to an expected number of $86.8 \pm 5.6$ events. The expected number of $\overline{\nu}_{e}$ interactions derives from the calculated flux of antineutrinos from the nuclear power plants.  Under the assumption of CPT invariance the observed deficit in the reactor $\overline{\nu}_{e}$ flux and the observed flavor change of solar $\nu$'s point to neutrino oscillation as a consistent explanation of all experimental data. An oscillation analysis of the available data yields good agreement between the oscillation parameters for $\nu$ and $\overline{\nu}$ \cite{hitoshi}.

With a livetime of 766.3 ton-yr KamLAND has recently published a more accurate measurement of the reactor $\overline{\nu}_{e}$ flux and spectrum providing unambiguous evidence from KamLAND alone for the oscillation of reactor antineutrinos. The shape of the energy spectrum measured by KamLAND is inconsistent with the energy spectrum of reactor $\overline{\nu}_{e}$ in the absence of oscillations at the 99.6\% C.L. For a constant baseline the neutrino survival probability $P_{ee} = 1- \sin^22\theta\sin^2\left(\Delta m^2 \frac{L}{E_{\nu}}\right) $ depends on the neutrino energy $E_{\nu}$, and spectral distortions are a characteristic signature of the oscillation effect. The current limits on neutrino oscillation parameters from reactor and solar experiments are summarized in Figure~\ref{alltan2004-2}.

\begin{figure}[tp]
\centerline{\epsfxsize=3.2in\epsfbox{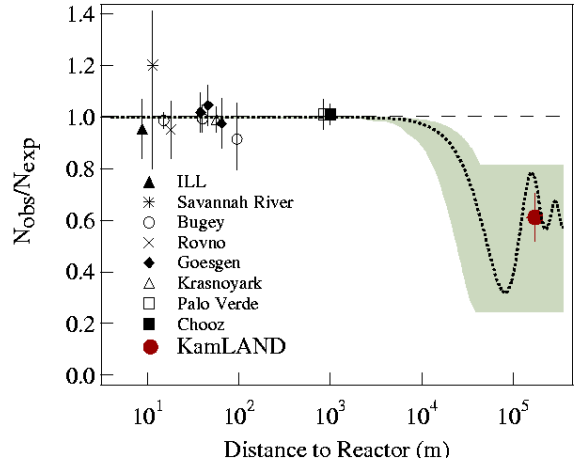}}   
\caption[]{Ratio of the $\overline{\nu}_{e}$ flux measured in reactor experiments to the expected  $\overline{\nu}_{e}$ flux in the absence of neutrino oscillation as a function of baseline. The shaded-region indicates the range of flux predictions corresponding to the 95\% C.L. large-mixing-angle region found in a global analysis of solar $\nu$ data. KamLAND made the first observation of the disappearance of $\overline{\nu}_{e}$ and confirmed the oscillation predictions from solar neutrino experiments. Figure from \cite{KamLAND1stPRL}. \label{inter}}
\end{figure}

\begin{figure}[tbp]
\centerline{\epsfxsize=3.5in\epsfbox{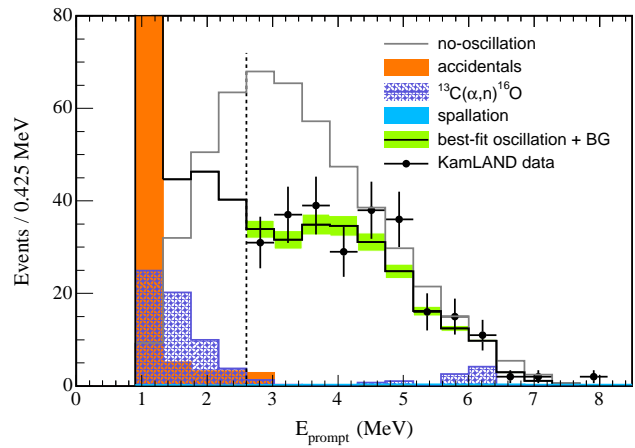}}   
\caption[]{Prompt energy spectrum of $\overline{\nu}_{e}$ candidate events and associated background spectra. The shaded region indicates the systematic error in the best-fit reactor spectrum above 2.6~MeV. The observed $\overline{\nu}_{e}$ spectrum is not only suppressed but incompatible with the expected spectrum at 99.6\% C.L. Figure from \cite{KamLAND2ndPRL}. \label{inter}}
\end{figure}

\section{Evidence for Neutrino Mass in Oscillation Experiments}

\begin{figure}[h]
\rightline{\epsfxsize=3.5in\epsfbox{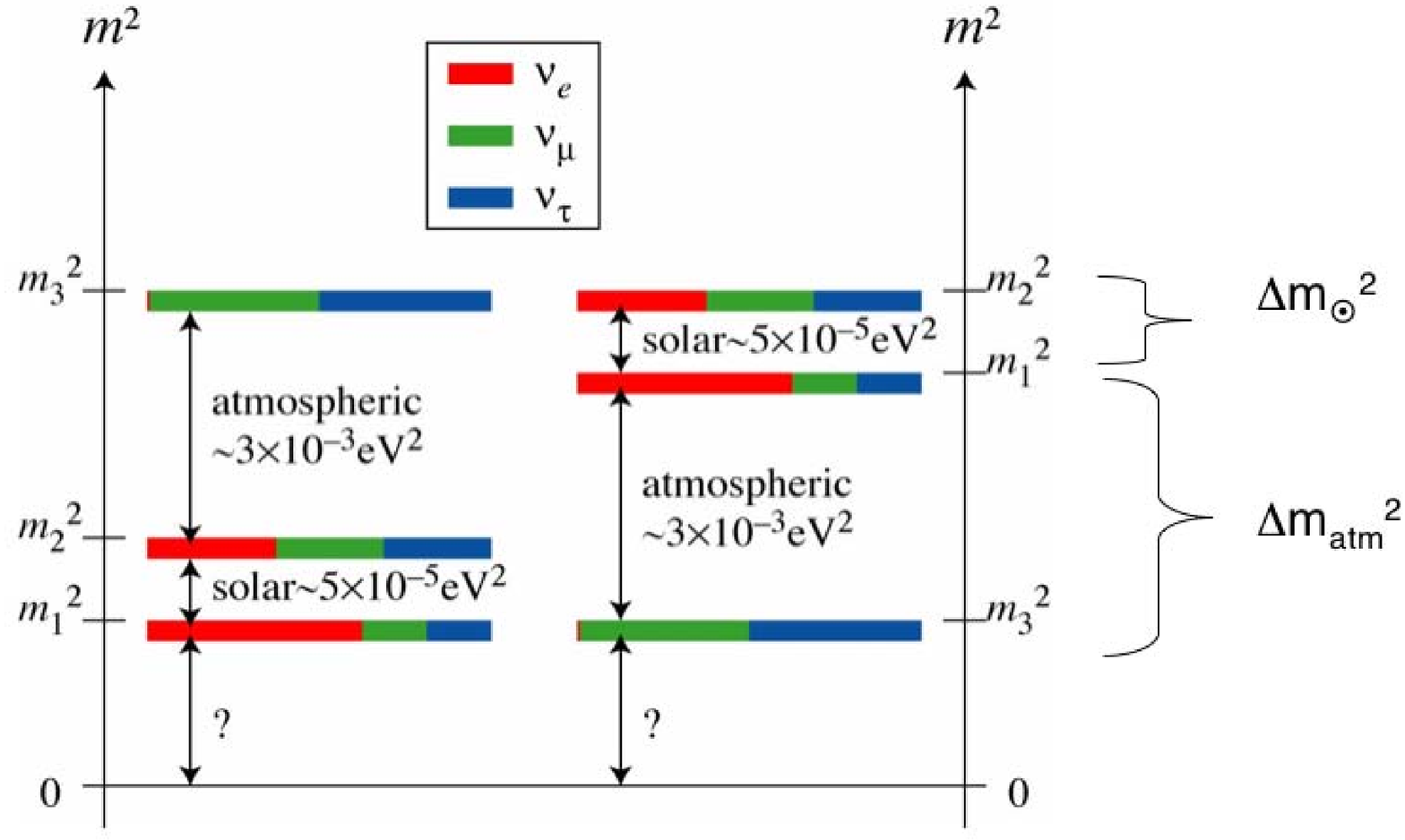}}   
\caption{Normal and inverted mass spectrum for three neutrino states. The mass differences have been measured precisely in oscillation experiments with various baselines using neutrino and antineutrino sources with different energies. \label{massspectrum}}
\end{figure}

\begin{figure}[h]
\centerline{\epsfxsize=2.2in\epsfbox{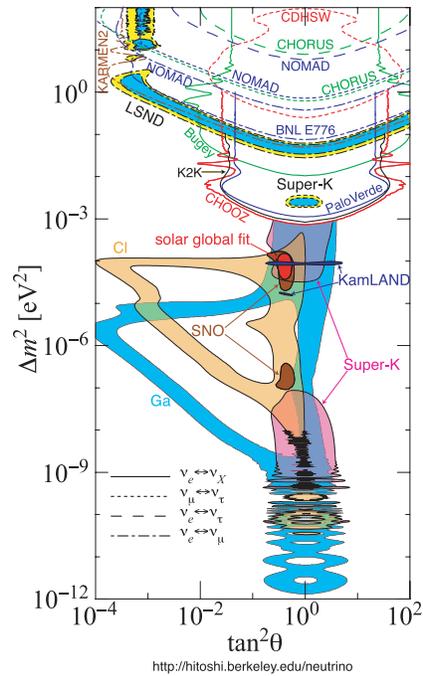}}   
\caption[]{Neutrino oscillation parameters as measured by atmospheric and accelerator experiments (``atmospheric region'') and solar neutrinos and reactor antineutrinos (``solar region''). The global fit of all solar experiments is consistent with the oscillation parameters of reactor antineutrinos under CPT invariance. Figure from \cite{hitoshi}. \label{alltan2004-2}}
\end{figure}

Experimental studies of terrestrial, atmospheric, and solar neutrinos  have established the flavor change and mixing of massive neutrinos. Measurements of atmospheric and accelerator experiments and solar and reactor neutrino observatories yield two different mass scales for the oscillation: $\Delta m^2_{atm} \sim 2.0 \times 10^{-3}~\rm{eV}^2$ and $\Delta m^2_{sol} \sim 7.1 \times 10^{-5}~\rm{eV}^2$. These measurements define the relative mass scale and allow  two possible mass spectra, as shown in Figure~\ref{massspectrum}. The absolute scale of the mass spectra is yet unknown but the minimum scale is given by the larger mass splitting $m \ge \sqrt{\Delta m^2_{atm}} \simeq 50 ~{\rm meV}$. The mixing angles associated with the atmospheric and solar transitions are nearly maximal and large, respectively. Combining the current results from all oscillation experiments we obtain the allowed $\Delta m^2$-$\tan^2\theta$ oscillation parameter regions shown in Figure~\ref{alltan2004-2}.

\clearpage

\section{Direct Neutrino Mass Measurements}

Over the past decades there has been steady progress in probing neutrino masses through direct measurements of decay kinematics. Direct kinematical measurements of neutrino masses give values consistent with zero. Techniques for measuring the mass of the electron neutrino involve the search for a distortion in the shape of the $\beta$-spectrum in the endpoint region. Tritium $\beta$-decay is commonly used for this measurement because of its low endpoint energy and simple nuclear and atomic structure. Tritium $\beta$-decay experiments use electromagnetic or magnetic spectrometers to analyze the momentum of the electrons and to infer the endpoint energy of the spectrum. The current best limits of $m_{\nu_{e}} \le 2.2$~eV/c$^2$ at 90\% C.L. comes from the Mainz and Troitsk neutrino mass experiments \cite{tritium}. A new experiment, the Karlsruhe Tritium Neutrino Experiment (KATRIN), with an expected sensitivity of 0.2~eV at 90\% C.L. is under construction \cite{KATRIN}.

\begin{figure}[bp]
\centerline{\epsfxsize=2.6in\epsfbox{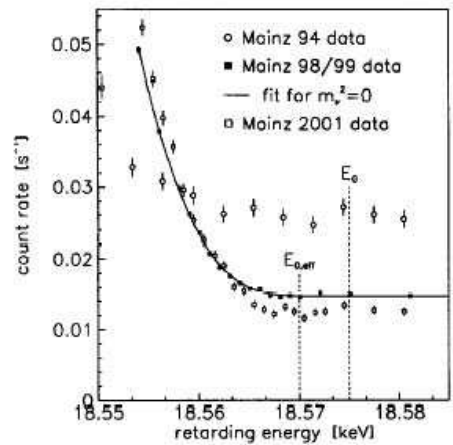}}   
\caption[]{Average count rate of tritium $\beta$-decays near the endpoint of 18.6~keV in the Mainz neutrino mass experiment. The count rate is shown as a function of the retarding  energy of the spectrometer. An analysis of this data yields an upper limit on the neutrino mass of $m_{\nu_{e}} \le 2.2$~eV/c$^2$ at 90\% C.L. Figure from \cite{tritium}. \label{inter}}
\end{figure}

Direct limits on both the muon and tau neutrino masses are based on kinematic measurements using semileptonic, weak particle decays.  The observables in these measurements are either invariant mass or the decay particle momentum. As these measurements rely on knowing the particle mass and momentum the sensitivity of these measurements to the neutrino mass is limited. Cosmology and nucleonsynthesis as well as the supernova 1987A set limits far lower than those placed by direct mass measurements. The direct experimental limits on neutrino mass as reported by the Particle Data Group \cite{PDG} are summarized in Table~\ref{numasslimits}.

Direct kinematic methods have not yet measured a non-zero neutrino mass. At present there is no direct indication from these experiments for new physics beyond the Standard Model and other searches for the signature of massive neutrinos are needed. 

\begin{table}[htb]
\caption[]{Experimental limits on neutrino masses from direct mass measurements.}
\begin{tabular}{llll}
\hline
Neutrino Mass & Mass Limit & Decay Mode & Experiment \\
\hline 
$m_{\nu_{e}}$ 		& $< 2.2~{\rm eV}$		& $^3\rm{H} \rightarrow$~$^3\rm{He} + e^- +\nu_{e}$	& Mainz \cite{tritium} \\
$m_{\nu_{\mu}}$	& $< 190~{\rm keV}$		& $\pi^+ \rightarrow \mu^+ + \nu_{\mu}$	&PSI  \cite{muondecay}  \\
$m_{\nu_{\tau}}$	& $< 18.2~{\rm MeV}$	& $\tau^- \rightarrow 2\pi^-\pi+\nu_{\tau}$ 	& ALEPH  \cite{ALEPH} \\
				&					& $\tau^- \rightarrow 3\pi^-2\pi+\nu_{\tau}$& \\
\hline
\end{tabular}
\label{numasslimits}
\vspace*{-13pt}
\end{table}

\section{Neutrino Constraints from Cosmology}

Stable neutrinos with masses as large as the limits from direct kinematic measurements would certainly overclose the Universe, i.e. contribute such a large cosmological density that the Universe could have never attained its present age. Cosmology implies a much lower upper limit on these neutrino masses. Considering the freezeout of neutrinos in  the early Universe it can be shown that the mass density and the sum of the neutrino masses are related as
\begin{eqnarray}
\sum m_{\nu_{x}} = 93 \Omega_{m} h^2~\rm{eV}
\end{eqnarray}
where $\Omega_{m}$ is the mass contribution to the cosmological constant. Analysis of the cosmic microwave background anisotropy combined with the galaxy redshift surveys and other data yield a constraint on the the sum of the neutrino masses of $\sum m_{\nu_{i}} \le 0.7~\rm{eV}$ \cite{WMAP}. The model dependence of this result is presently under discussion. Big Bang nucleonsynthesis constrains the parameters of possible sterile neutrinos which do not interact and are produced only by mixing. The current limit on the total number of neutrinos from Big Bang nucleosynthesis is $1.7 \le N_{\nu} \le 4.3$ at 95\% C.L.

\section{Probing the Nature of Neutrinos and $\nu$ Mass in 0$\nu\beta\beta$}

Another unique signature of massive neutrinos is neutrinoless double $\beta$-decay, a lepton-number-violating process also known as 0$\nu\beta\beta$. The process $(A, Z) \rightarrow (A, Z+2) +2e^-$ can be mediated by an exchange of a light Majorana particle, or an exchange of other articles. The existence of  0$\nu\beta\beta$ requires the existence of Majorana neutrino mass independent of the mechanism of the process. Neutrinoless double-beta decay is the only experimental approach known to date that distinguishes between Majorana and Dirac masses.  The experimental signature of  0$\nu\beta\beta$ is a peak in the combined electron spectrum at the $Q_{\beta\beta}$-value of the reaction. The  observable  0$\nu\beta\beta$-decay rate $1/T^{0\nu}_{1/2}$ is proportional to the effective Majorana mass squared $\left|\left<m_{\beta\beta}\right>\right|^2$
\begin{eqnarray}
1/T^{0\nu}_{1/2} = G{^0\nu} \left| M^{0\nu}\right|^2 \left|\left<m_{\beta\beta}\right>\right|^2
\end{eqnarray}
with $\left<m_{\beta\beta}\right> = \sum_{i}U^2_{ei}m_{\nu i}$. The lifetime measurement is translated into an effective Majorana mass using nuclear structure calculations which in turn can be used to set upper limits on the neutrino mass. 
The phase factor $G^{0\nu}$ can be calculated reliably but there is significant uncertainty in the calculations of the matrix elements $M^{0\nu}$.

The best current limits on $T^{0\nu}$ and $\left<m_{\beta\beta}\right>$ come from the Heidelberg-Moscow experiment which used 11~kg of enriched $^{76}$Ge with an isotopic abundance of 86\% \cite{klapdor2001}. Until recently, the Heidelberg-Moscow collaboration reported a lower limit on the half-life and upper limit on the effective neutrino mass:
\begin{eqnarray*}
T^{0\nu}_{1/2} \ge 1.9 \times 10^{25} ~{\rm yr~(90\% ~C.L.)} \\
m_{\beta\beta} \le 0.35 ~{\rm eV~(90\%~C.L.)}
\end{eqnarray*}

\begin{figure}[tbp]
\centerline{\epsfxsize=2.5in\epsfbox{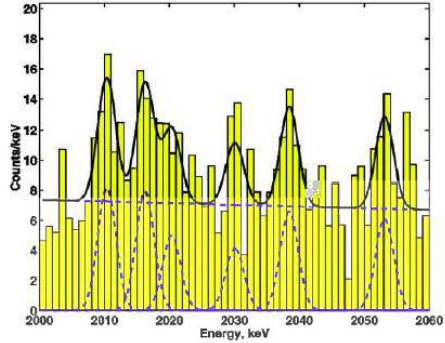}}   
\caption[]{Energy spectrum of 10.96 kg enriched $^{76}$Ge in the range of 2000-2060~keV. The line indicates the identified peaks including Bi backgrounds at 2010.7, 2016.7, 2021.8, 2052.9~keV and an additional signal at $\sim$2039~keV. This corresponds to the $Q$ value of the 0$\nu\beta\beta$ process. Figure from \cite{klapdor2004}. \label{inter}}
\end{figure}

A recent analysis of data from this experiment by Klapdor-Kleingrothaus {\em et al.} led to the announcement of the discovery of neutrinoless double-beta decay. Klapdor-Kleingrothaus {\em et al.} report a 4.2 $\sigma$ evidence for  0$\nu\beta\beta$  based on 71.7~kg-yr of data taken between August 1990-May 2003  \cite{klapdor2004}. These claims have not yet been confirmed.

%
%
%
%

\end{document}